# Heterostructure films of SiO$_2$ and HfO$_2$ for high power laser optics prepared by plasma-enhanced atomic layer deposition

Shawon Alam[1,2], Pallabi Paul[1,2], Vivek Beladiya[1,2], Olaf Stenzel[2], Marcus Trost[2], Steffen Wilbrandt[2], Sven Schröder[2], Gabor Matthäus[1], Stefan Nolte[1,2], Sebastian Riese[3], Felix Otto[4], Torsten Fritz[4], Alexander Gottwald[5], and Adriana Szeghalmi[1,2,*]

[1] Institute of Applied Physics, Friedrich Schiller University Jena, Albert-Einstein-Str. 15, 07745 Jena, Germany
[2] Fraunhofer Institute for Applied Optics and Precision Engineering, Albert-Einstein-Str. 7, 07745 Jena, Germany
[3] Layertec GmbH, Ernst-Abbe-Weg 1, 99441 Mellingen, Germany
[4] Institute of Solid State Physics, Friedrich Schiller University Jena, Helmholtzweg 5, 07743 Jena, Germany
[5] Physikalisch-Technische Bundesanstalt, Abbestraße 2-12, 10587 Berlin, Germany
* Correspondence: adriana.szeghalmi@iof.fraunhofer.de; Tel.: 0049807320

**Abstract:** Absorption losses and laser-induced damage threshold (LIDT) are considered as the major constraint for the development of optical coatings for high-power laser optics. Such coatings require paramount properties like low losses due to optical absorption, high mechanical stability, and enhanced damage resistance to withstand high-intensity laser pulses. In this work, heterostructure films were developed by the intermixing of SiO$_2$ and HfO$_2$ using plasma-enhanced atomic layer deposition (PEALD) technique. Thin film characterization techniques such as spectroscopic ellipsometry, spectrophotometry, substrate curvature measurements, x-ray reflectivity, and Fourier transform infrared spectroscopy were employed for extracting optical constants, spectral interpretation, residual stress, layer formation, and functional groups present in the heterostructures, respectively. These heterostructures demonstrate tunable refractive index, bandgap, and improved optical losses and LIDT properties. The films were incorporated into antireflection coatings (multilayer stacks and graded index coatings) and the LIDT was determined at 355 nm wavelength by the R-on-1 method. Optical absorptions at the reported wavelengths were characterized using photothermal common-path interferometry and laser-induced deflection techniques.

## 1. Introduction

The research on fabricating optical coatings for high power laser optics in the ultraviolet and near-infrared spectral range has been a center of attention for a long time due to the scarcity of thin film materials that are transparent or experience low absorption and exhibit high damage resistance in such spectral ranges [1,2]. The laser-induced damage threshold (LIDT) is a major constraint for developing such coatings for high-power laser systems as the optics experience damages upon high-intensity laser irradiation [3–6]. Several oxides such as TiO$_2$, Nb$_2$O$_5$, ZrO$_2$, Ta$_2$O$_5$, HfO$_2$, Sc$_2$O$_3$, Al$_2$O$_3$, SiO$_2$, and various fluorides are prevalent materials for multilayer optical coatings comprised of alternating low and high refractive index material [7–10]. Among these materials, HfO$_2$ is a promising high refractive index substance that attracted much attention on account of its exclusive properties such as a wide bandgap (5.5-5.7 eV), thermal stability, high laser damage resistance, and a broad transparency range from ultraviolet to mid-infrared (0.22-12 μm) [11–15]. HfO$_2$ films are preferred materials to replace SiO$_2$ as gate dielectric oxides in semiconductor applications due to their high dielectric constant (25-28), density, and ductility [12,16–20]. SiO$_2$ is known as a prominent low refractive index material used for high power laser coatings from the ultraviolet to the infrared spectral range because of its broad transparency range (0.2-8 μm) [21–23]. Such films exhibit low losses due to optical absorption, dielectric constant (3.9), and large optical bandgap (8.9-9.0 eV) properties [14,18,24–26]. SiO$_2$ finds applications in microelectronic devices as a gate insulator and development of photovoltaics [17,25,27–30]. Most of the materials have an optical absorption near the ultraviolet spectral region due to the limited width of the bandgap [1,7]. Optical coatings consisting of single oxide films are limited to a few applications only and require materials with intermediate refractive indices (not obtainable from single oxides) for broadening the application window. Finding materials with suitable optical characteristics (high LIDT and low absorption losses) for high power laser coatings are the key challenge because such materials in the UV spectral range are limited.

Atomically mixing low and high index materials from these sets of oxides to realize heterostructures provides new capabilities to produce thin films with tunable refractive index, residual stress, and optical bandgap, which can expand the application window [31–36]. HfO$_2$ as a high index material is usually combined with the low refractive index oxide SiO$_2$ to form multilayers coating that can exhibit intermediate refractive indices between those of the constituents and achieve the reduction of optical losses due to the extension of transparent ranges down to the deep UV spectral range [14,26,37]. Heterostructures can also tune laser damage resistance due to the modified optical bandgap values as reported by Gallais et. al [8]. Among other heterostructures (high and low index combination), SiO$_2$:HfO$_2$ is selected for optical coatings because of their thermomechanical stability and high LIDT [1,38–40].

Graded-index coatings have been observed to achieve enhanced film properties and laser damage resistance due to the reduction of discontinuous interfaces [34,41]. Such coatings are also called rugate or inhomogeneous coatings that consist of continuously varying refractive index, n along with film thickness rather than abrupt change of n as in the discrete multilayer stacks [41–43]. However, the varying refractive index can be obtained by adjusting the composition of the heterostructures constituent materials at the atomic level [9,44]. The graded-index coatings can substantially minimize the reflection harmonics by matching the intermediate refractive indices between the substrate and the air medium [26,45,46].

The amorphous state of the thin film is desired in optical coatings because crystalline films generate grain boundaries that lead to unwanted losses due to scattering and absorption [26,47,48]. HfO$_2$ films tend to crystallize at a relatively low deposition temperature as reported by several authors [49,50]. The previous work from our group found a thickness and temperature-dependent crystallinity of HfO$_2$ [51]. The heterostructures can also mitigate crystallization by incorporating a thin layer of an amorphous film within such coatings [44,48,52]. The inclusion of SiO$_2$ in HfO$_2$ films can significantly increase the crystallization temperatures of HfO$_2$ to much higher ranges which means the film can retain the amorphous state even grown at a high temperature [26,53].

The deposition of heterostructure coatings is quite challenging because the film requires versatile properties such as low absorption losses, enhanced mechanical stability, and high damage resistance to withstand intense laser irradiation. Traditional methods such as electron beam evaporation (EBE), ion-beam sputtering (IBS), ion-assisted deposition (IAD), sol-gel method, and chemical vapor deposition (CVD) are applied for high-power laser optical coatings [5,12,15,47]. However, due to the downscaling of device sizes and 3D photonic integration, it is essential to ensure atomic-scale precision along with 3D conformality on arbitrary substrate geometries. Atomic layer deposition (ALD) is a powerful technique for preparing optical coatings for high-power laser optics that allows deposition at a lower temperature using the variant plasma-enhanced ALD [1,25,54,55]. ALD is based on two self-limiting surface reactions that provide atomic-scale control over the film thicknesses and conformity over complex-shaped substrates, which are suitable for high power laser components such as aspheric lenses, laser mirrors, beam expanders, polarizers, and beam splitters for industrial applications [12,16,31,56,57]. ALD grown films are homogeneous, smooth, highly dense, and pinhole-free which exhibits high resistance to laser-induced damage [1,47,58,59].

This work intends to provide a comprehensive understanding of $SiO_2$:$HfO_2$ heterostructure films prepared by the plasma-enhanced atomic layer deposition (PEALD) process. The heterostructure films were analyzed by employing several state-of-the-art characterization techniques to understand the tunability of various properties as well as realizing coatings for high power laser applications.

## 2. Materials and Methods

2.1 Deposition of $SiO_2$:$HfO_2$ thin film heterostructures

Single-layer thin films of $SiO_2$, $HfO_2$, and $SiO_2$:$HfO_2$ heterostructures (varying the composition ratio of $SiO_2$ and $HfO_2$) were fabricated by using a SILAYO-ICP330 PEALD vacuum reactor (Sentech Instruments GmbH, Berlin, Germany) [12,60]. Such instruments incorporate an inductively coupled plasma source (RF generator, 13.56 MHz, up to 500 W) with the possibility of implementing both thermal and plasma-enhanced ALD processes. Thin films of $SiO_2$ and $HfO_2$ were prepared using the bis[diethylamino]silane (BDEAS) and tetrakis-dimethylamino hafnium (TDMAH) precursors via PEALD processes. Such films were grown on fused silica (diameter 25 mm, thickness 1 mm) and silicon wafer substrates to investigate optical properties by employing several characterization techniques. The film growth by PEALD depends on several process parameters like growth temperature, precursors pulse-purge duration, and the flow rate of carrier gases that influences the optical and mechanical properties of the film [61,62]. Such parameters for $SiO_2$ and $HfO_2$ films were previously optimized in our works which can facilitate reproducible film growth per cycle and uniformity in thicknesses within the ALD framework [51,60]. Therefore, the optimized process parameters were incorporated into the deposition of heterostructure films. The process parameters of the PEALD thin film depositions are listed in table 1.

Table 1. PEALD process parameters for $SiO_2$:$HfO_2$ heterostructure films.

| Thin films | Precursor Pulse\|Purge (s) | Co-reactant Pulse\|Purge (s) | $O_2$ flow (sccm) | Ar flow (sccm) |
|---|---|---|---|---|
| $SiO_2$ | BDEAS: $SiH_2[N(CH_2CH_3)_2]_2$ 0.3\|5 | $O_2$ Plasma 3\|2 | 200 | 30 |
| $HfO_2$ | TDMAH: $Hf[N(CH_3)_2]_4$ 3\|5 | $O_2$ Plasma 5\|5 | 200 | 160 |

sccm: Standard cubic centimeter per minute

The deposition temperature was kept at 100°C for all the processes within the framework of this work to suppress temperature-dependent crystallization of $HfO_2$ thin films. The plasma power was kept at 100 W. The precursor exposure and separation step for the growth of $SiO_2$ and $HfO_2$ varied due to the vapor pressure of the precursor and physisorption energy that limits the purge duration [30,44]. Argon was utilized as both precursor carrier and purge gas for these films. The purge steps are applied to prevent gas-phase reactions between the precursors (reactant) and co-reactant (oxidant agent) in the reactor [30,44,63].

*2.2 Characterizations of the SiO$_2$: HfO$_2$ heterostructure films*

Several analytical techniques have been applied to characterize the PEALD thin films for extracting optical constants (refractive index and extinction coefficient), film thickness, spectral performance (reflectance and transmittance), mechanical stress, surface roughness, composition and the laser-induced damage threshold. Optical constants and thickness of the films grown on Si (100) substrates were estimated by spectroscopic ellipsometry (Sentech Instruments GmbH, Berlin, Germany) measurements ($\Psi$ and $\Delta$ spectra due to the changes in amplitude and phase of incident light upon reflection). A Tauc-Lorentz oscillator model-based approach was implemented to determine the optical constants of the heterostructures by fitting the measured ellipsometric parameters. Further, the indirect optical bandgap of such structures is estimated using the Tauc-plot calculations.

The reflectance R and transmittance T spectra were measured using a spectrophotometer device [Lambda 950, PerkinElmer Inc., USA] from 190 to 1200 nm wavelength within a measurement accuracy of ±0.3%. The optical losses are calculated using the following expression [64–66].

$$\text{Optical losses \%} = 1-T-R$$

In order to extract the optical constants of the heterostructures in the vacuum ultraviolet (VUV) spectral ranges, further T/R measurements were carried out between 125 - 225 nm at 2° angle of incidence at the Physikalisch Technische Bundesanstalt, Berlin [67,68]. The measurement uncertainties in the VUV range were between 0.8% and 1.2%. The optical constants were further extracted from the VUV transmittance and reflectance spectra using a Lorentzian calculator (LCalc) [64].

Controlling the mechanical stress in thin films is crucial for their longevity and performance. The mechanical stress of the heterostructures was estimated using an FLX-2320 instrument (KLA-Tencor GmbH, Dresden, Germany) which incorporates the measurement of change in radius of curvature of the substrate before and after the deposition of the PEALD films. The stress values in the films are calculated from these substrate curvature measurements using the Stoney equation [69–71].

$$\sigma = \frac{1}{6} \cdot \frac{E_s}{(1-\nu_s)} \cdot \frac{t_s^2}{t_f} \left(\frac{1}{R_f} - \frac{1}{R_s}\right)$$

where $\sigma$ is the mechanical stress, $t_s$ and $t_f$ are substrate and film thickness, $R_s$ and $R_f$ are the radius of curvature of the bare (substrate without film before deposition) and coated wafer (substrate with film after deposition). $E_s$ and $\nu_s$ indicates Young's modulus and Poisson's ratio of the substrate.

X-ray reflectivity (XRR) measurements were conducted on a Si wafer to estimate the thickness, surface roughness, and layer density of the heterostructure films [30,72]. The XRR instruments (Bruker AXS, Karlsruhe, Germany) consist of a monochromatic X-ray beam (Cu-K$\alpha$ radiation at $\lambda$ = 0.154 nm) for such measurements at grazing incident angle from 0° to 8° [73]. The measured spectra were fitted using the Brucker Leptos 7 software for obtaining the film properties [72].

Fourier transform infrared spectroscopy (FTIR) (Varian Inc., Palo Alto, CA, USA) for heterostructures on Si wafers was performed to investigate functional groups present on the film material [63,74]. This technique utilized mid-infrared radiation (from 400 to 4000 $cm^{-1}$) exciting atomic or molecular vibrations in that wavelength region [63,75]. However, the measurements require a baseline correction for absorption peak extraction using data analysis software.

AFM measurements were conducted on selected heterostructures to further investigate their surface roughness. A Dimension 3100 (Bruker, Billerica, MA, USA) equipped with a Nanoscope controller (Digital Instruments, Tonawanda, NY, USA) was used at ambient conditions. The used Si-tips had a typical radius of less than 10 nm. Sample areas of 1 × 1 μm$^2$ and 2 × 2 μm$^2$ were scanned in tapping mode, and their surface roughness was calculated in root-mean-square (rms).

Laser-induced damage threshold (LIDT) tests of the heterostructure films on fused silica substrates were performed using high-intensity laser irradiation of nanosecond pulses. The ISO-21254 standard framework defines LIDT as the maximum quantity of laser fluence or energy density an optics can withstand for which the extrapolated probability of damage is zero [76,77]. The R-on-1 test method was deployed using a Nd:YAG laser, frequency-tripled, Q-switched LITRON NanoTRL-650-10 [Litron Lasers, Rugby, England] in order to estimate the laser induced damage threshold of the coatings. The effects of linear absorption on the LIDT of the heterostructure films were estimated by employing the photothermal common-path interferometry (PCI) method. These measurements were carried out on the fused silica substrates at 355 nm wavelength.

## 3. Results

Single-layer thin films of SiO$_2$ and HfO$_2$ on Si wafers were prepared by the PEALD processes before proceeding towards heterostructures. Spectroscopic ellipsometry was employed to determine thicknesses for such films. The growth per cycle (GPC) of the films was estimated through dividing the thicknesses by the number of cycles applied. HfO$_2$ films exhibit relatively higher GPC (1.70 Å/cycle) compared to SiO$_2$ (1.17 Å/cycle) [12,56,57,78]. The refractive indices of SiO$_2$ and HfO$_2$ are reported at 355 and 1064 nm wavelengths for the optical designs carried out in this work. High index HfO$_2$ film exhibits 2.01 at 355 nm and 1.93 at 1064 nm wavelengths whereas SiO$_2$ marked 1.47 at 355 nm and 1.45 at 1064 nm wavelengths as the refractive indices. These results are consistent with the refractive indices with a little deviation (due to different deposition methods and process parameters) reported by other authors [50,79,80]. The knowledge obtained from the single-layer thin films was applied to the deposition of heterostructures.

The heterostructures were prepared from the alternating layers of SiO$_2$ and HfO$_2$ films with a specific number of bilayers to realize a half-wave optical thickness at 1064 nm wavelength [40,81–83]. However, ultrathin layers of the heterostructures (approximately 25 nm) were grown at the beginning to estimate the GPC and refractive indices at the target wavelength. The composition, total thickness, and refractive indices of a series of heterostructures deposited on the Si wafer and fused silica substrates are shown in table 2. The heterostructure films were categorized into several groups with gradually increasing cycles of HfO$_2$ for 2, 4, 8, 16, and 64 cycles of SiO$_2$.

Table 2. The composition, total thickness, and refractive indices of SiO$_2$:HfO$_2$ heterostructures.

| Sample | ALD cycles of SiO$_2$ | ALD cycles of HfO$_2$ | Cycle ratio SiO$_2$:HfO$_2$ | Number of Bilayers | Thickness [nm]±0.1 | Refractive index [355 nm] ±0.01 | Refractive index [1064 nm] ±0.01 |
|---|---|---|---|---|---|---|---|
| **S1** | 2 | 2 | 1:1 | 615 | 307.1 | 1.83 | 1.76 |
| S2 | 2 | 8 | 1:4 | 100 | 149.5 | 1.99 | 1.90 |

| S3 | 2 | 16 | 1:8 | 98 | 293.3 | 2.01 | 1.92 |
| --- | --- | --- | --- | --- | --- | --- | --- |
| S4 | 4 | 2 | 2:1 | 465 | 325.1 | 1.71 | 1.66 |
| **S5** | 4 | 4 | 1:1 | 315 | 323.6 | 1.82 | 1.75 |
| S6 | 4 | 8 | 1:2 | 160 | 275.8 | 1.91 | 1.83 |
| S7 | 4 | 16 | 1:4 | 87 | 271.1 | 1.97 | 1.88 |
| S8 | 8 | 2 | 4:1 | 300 | 342.3 | 1.62 | 1.58 |
| S9 | 8 | 4 | 2:1 | 230 | 324.8 | 1.73 | 1.67 |
| **S10** | 8 | 8 | 1:1 | 145 | 301.3 | 1.83 | 1.76 |
| S11 | 8 | 16 | 1:2 | 76 | 260.6 | 1.93 | 1.84 |
| **S12** | 16 | 16 | 1:1 | 72 | 301.5 | 1.84 | 1.77 |
| S13 | 64 | 2 | 32:1 | 20 | 160.5 | 1.50 | 1.47 |
| S14 | 64 | 4 | 16:1 | 20 | 166.7 | 1.52 | 1.49 |
| S15 | 64 | 8 | 8:1 | 20 | 180.7 | 1.57 | 1.53 |
| S16 | 64 | 16 | 4:1 | 20 | 200.4 | 1.66 | 1.61 |

For example, the first group incorporates sample S1 to S3 that comprises 2 cycles of $SiO_2$ with 2, 8, and 16 cycles of $HfO_2$. Several heterostructures (S1, S5, S10, and S12) were emboldened in the table as they belong to identical cycle ratio groups with different numbers of cycles of the constituents. The refractive indices of the heterostructures varied at the reported wavelengths (355nm and 1064 nm) due to their composition and stayed between $SiO_2$ and $HfO_2$. The film thickness and optical constants were extracted through the Tauc-Lorentz model upon fitting the ellipsometry spectra ($\Psi$ and $\Delta$, generated at a 70º angle of incidence from 200 to 980 nm wavelength in ambient environment). The oscillator dispersion model consists of several fit parameters such as amplitude, center energy, broadening, and optical bandgap. The representative of heterostructures with 4 and 8 cycles of $SiO_2$ are shown in figure 1(a-d). The curves (measured and simulated) were in very good agreement for $\Psi$ and $\Delta$. The $\Psi$ and $\Delta$ for each heterostructure were slightly shifted spectrally with respect to each other due to the composition-dependent optical properties. Therefore, the Tauc-Lorentz dispersion relation demonstrated quite good results for estimating optical constants and film thicknesses of the heterostructures.

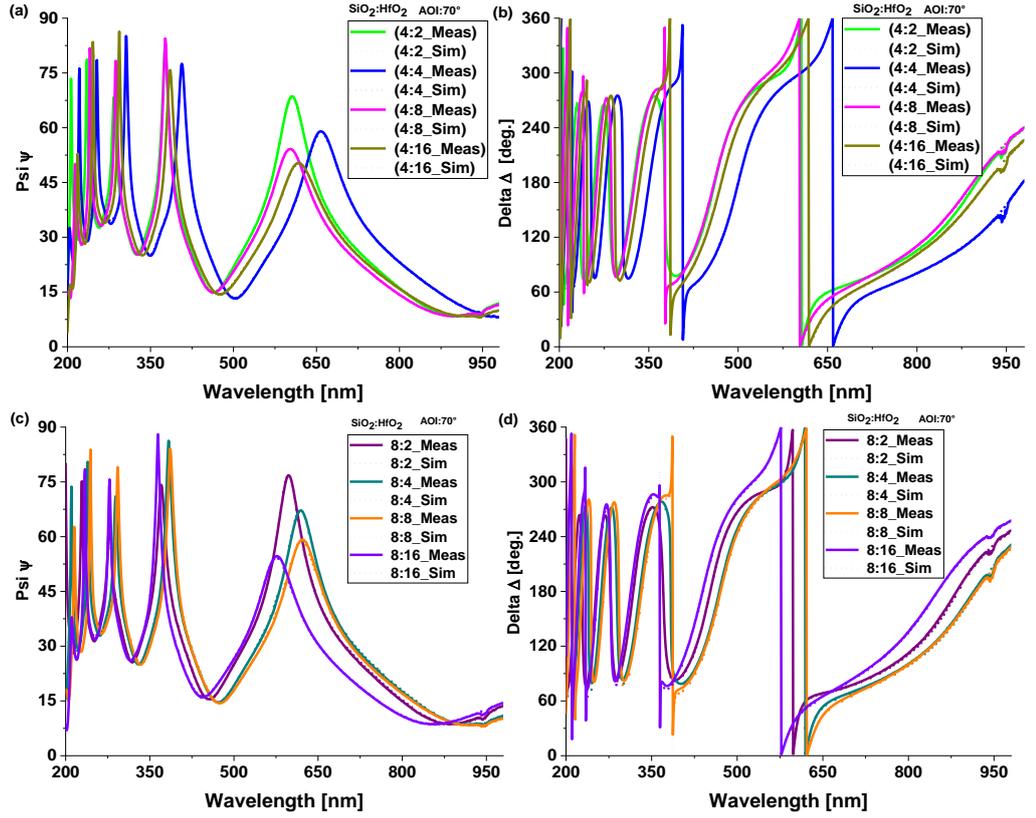

**Fig. 1.** Measured and simulated spectroscopic ellipsometric parameters (Ψ and Δ) for the $SiO_2$:$HfO_2$ heterostructures with (a, b) 4 and (c, d) 8 cycles of $SiO_2$.

The composition-dependent optical constants (n, k) obtained through a Tauc-Lorentz dispersion relation for $SiO_2$:$HfO_2$ heterostructures are included in figure 2(a-f) where the curves were reported between 200 and 1100 nm wavelength. It is evident from figure 2 (a-c) that the refractive indices curves remain between $HfO_2$ and $SiO_2$ meaning the intermediate refractive indices are obtainable from the heterostructures by atomically adjusting the constituents. These intermediate indices are not achievable from the single materials as no films exist with such refractive indices in nature, providing flexibility in the thin film coatings design which broadens the application window. In figure 2(b,c), the refractive indices curves of ultrathin layers of the heterostructures with 4 and 8 cycles of $SiO_2$ are also included along with thick films. The refractive index curves of the thicker films slightly shift compared to the thinner one due to the asymmetrical growth (related to density) and coatings formations at the edges (approximately 1 mm surface area) of the backside of the substrates. The refractive index values of the heterostructures corresponding to figure 2(a-c) are reported in table 2 for the two specific wavelengths, e.g., 355 nm and 1064 nm, respectively. The extinction coefficient curves of the heterostructures in figure 2(d-f) show that the absorption edge of $HfO_2$ is located nearly at 230 nm wavelength. However, the absorption edges for $SiO_2$ along with some heterostructures were not feasible to determine due to instrument limitation. The absorption edge is gradually shifting towards shorter wavelengths because of the heterostructures composition.

The dispersion profile of the heterostructures with identical cycle ratio are illustrated in Figure 3. They exhibit reasonably similar refractive indices as shown in Figure 3(a). The extinction coefficient profiles in Figure 3(b) indicate some variation among the heterostructures having similar composition ratio. This feature can be attributed to the change in stoichiometry of $HfO_2$ layer. As $SiO_2$ films become extremely thin, they tend to donate oxygen which could be diffused to the adjacent $HfO_2$ layer while compensating for the stoichiometry and reducing the extinction values. However, the number of interfaces in the heterostructures does not seem to have influence over the extinction spectra. The absorption edge of HfO₂ is located nearly at 230 nm wavelength. The slight blue shift of extinction coefficients upon decreasing the number of ALD cycles of SiO₂ and HfO₂ while keeping the cycle ratio unchanged might be an indication of the quantum size effect. Such effects were previously reported for Al₂O₃:TiO₂ heterostructures in our previous work [84].

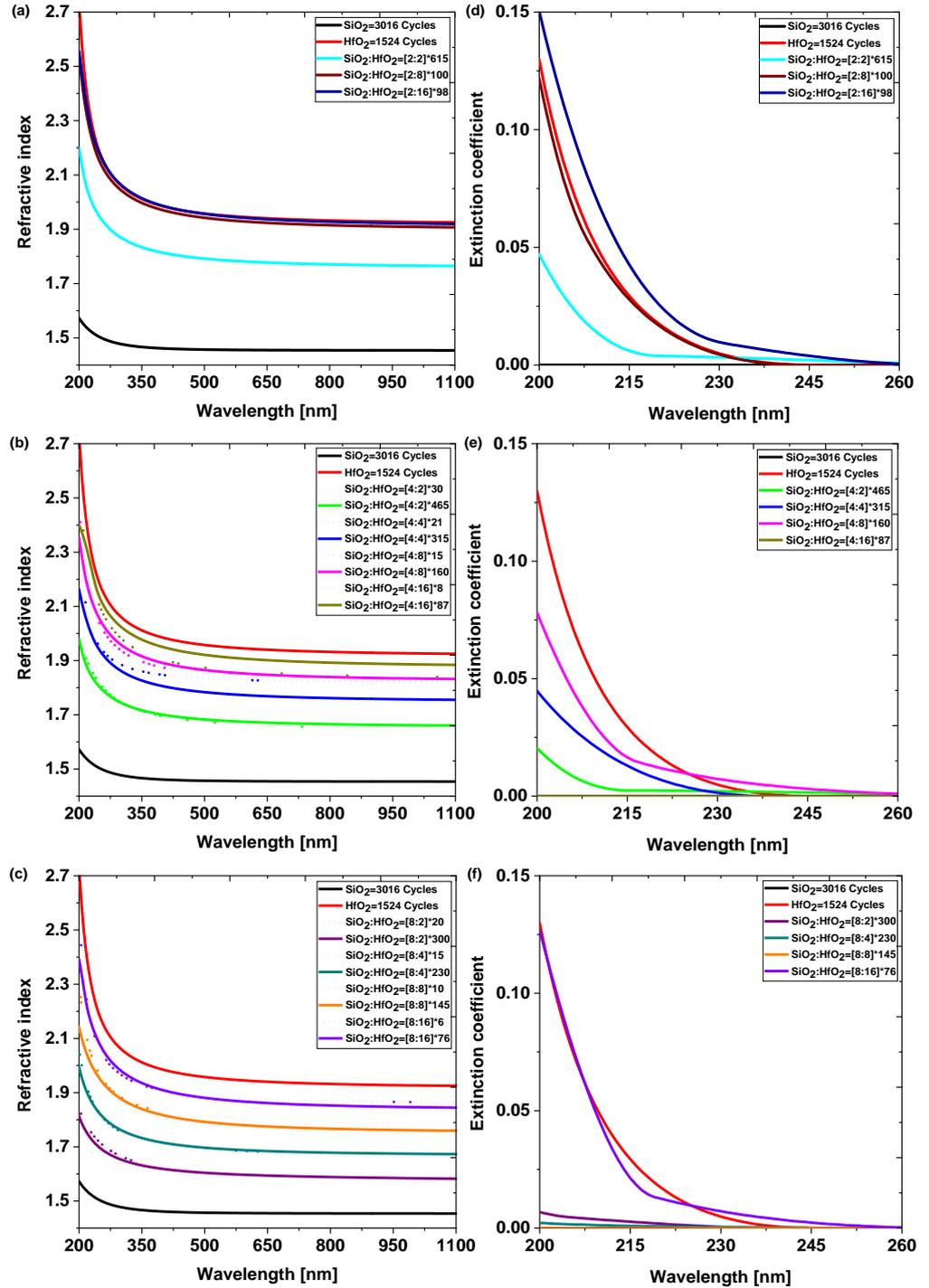

**Fig. 2.** Optical constants curves for the heterostructures with (a,b) 2, (c,d) 4, and (e,f) 8 cycles of SiO₂.

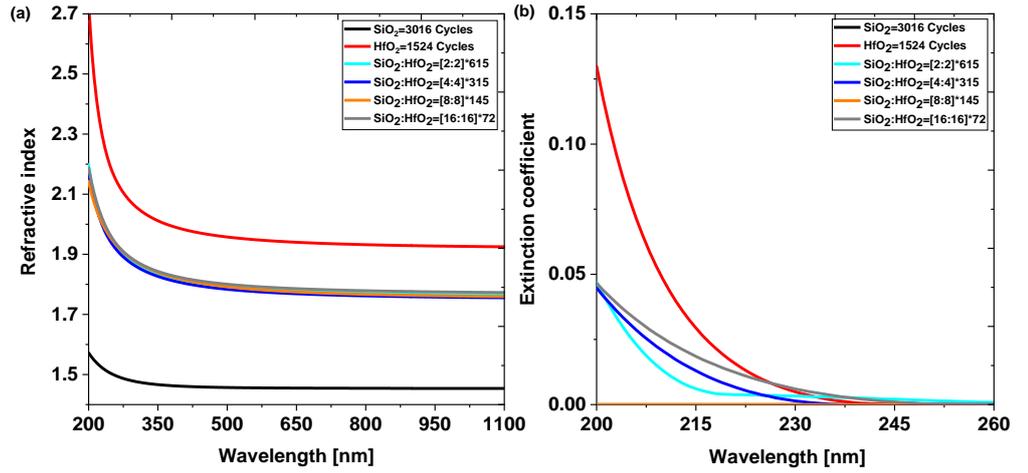

**Fig. 3**. Optical constants curves for the heterostructures with identical cycle ratios of the constituents.

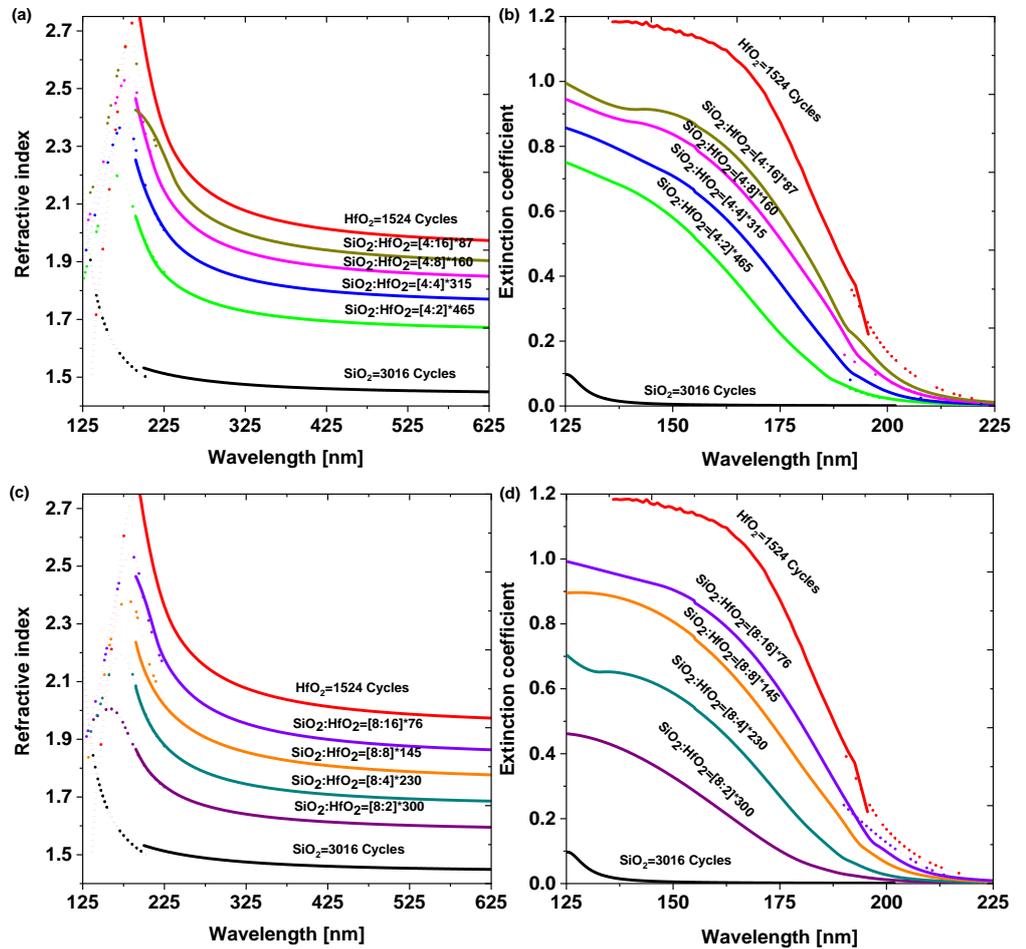

**Fig. 4.** Comparison of optical constants as a function of wavelength estimated from VUV and UV-VIS measurements for (a,b) 4 and (c,d) 8 cycles of SiO$_2$.

In order to estimate the optical constants of the heterostructures in the vacuum ultraviolet (VUV) ranges from 125 to 225 nm wavelength, the heterostructures with 4 and 8 cycles of SiO$_2$ were selected for the measurements. The optical constants extracted from VUV together with UV-VIS ranges are shown in figure 4 (a-d). The spectra obtained from the VUV measurements (dotted curves) are consistent with the UV-VIS observed trend (solid curves). However, slight spectral shifts visible in their intersecting ranges (190 - 220 nm wavelength) could be attributed to vacuum to air shift, or measurement and fitting uncertainties.

The optical bandgap of SiO$_2$ and HfO$_2$ were estimated by the Tauc plot calculations to be at 8.9 eV and 5.6 eV, respectively. These values are also consistent with the other literature, however slightly deviating depending on different deposition methods [11,15,25]. The bandgap of the heterostructures with 2, 4, and 8 cycles of SiO$_2$ were also estimated and graphically represented with refractive indices at 355 nm wavelength in figure 5. The optical bandgap values follow an inverse relationship with the refractive index. The bandgaps of the heterostructures are limited by HfO$_2$ since the inclusion of more SiO$_2$ content in the heterostructures enhances the bandgap. The heterostructures with identical cycle ratios e.g., [2:2]*615, [4:4]*315, and [8:8]*145, exhibit similar values of refractive indices within their determination accuracy, however, their optical bandgap can be tailored to a certain extent (for instance, here, 5.6 – 5.9 eV) depending on their individual layer thicknesses.

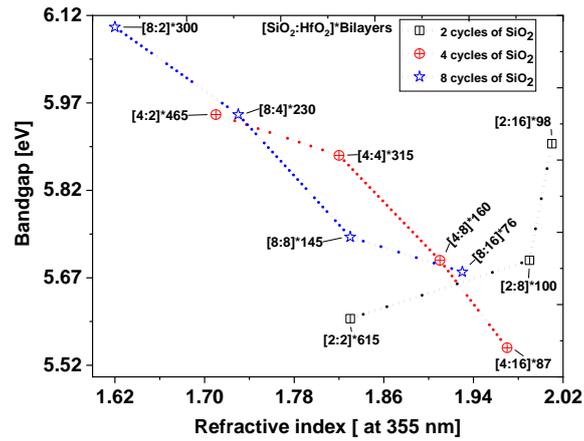

**Fig. 5.** Optical bandgap as a function of refractive index at 355 nm wavelength.

HfO$_2$ is transparent down to approximately 250 nm wavelength while SiO$_2$ shows transparency for a broad spectral range between deep UV and NIR spectral range. Creating the heterostructures enables enhanced optical properties such as extended transparency range depending on the composition of the two constituents. Figure 6 displays the transmission T/ reflection R and optical losses spectra of the SiO$_2$:HfO$_2$ heterostructures with identical cycle ratios. The deviations in transmission maxima and reflection minima for the compositions in Figure 6 (a) can be attributed to the difference in their total physical thickness. In Figure 6(b), single layer HfO$_2$ film exhibits higher losses in the UV spectral range due to its optical absorption edge. Therefore, this optical absorption edge can also be tailored by atomically controlling thickness of HfO$_2$ films through precise tuning of composition enabled by PEALD. Optical losses curves in Figure 6(b) indicate the shift of absorption edges towards shorter wavelengths due to the decrease in interlayer thickness (i.e., decreasing PEALD cycles of SiO$_2$ and HfO$_2$ while keeping the cycle ratios constant) attributed to the quantum size effects. Such trends are also consistent with the extinction coefficient curves as shown in Figure 3(b).

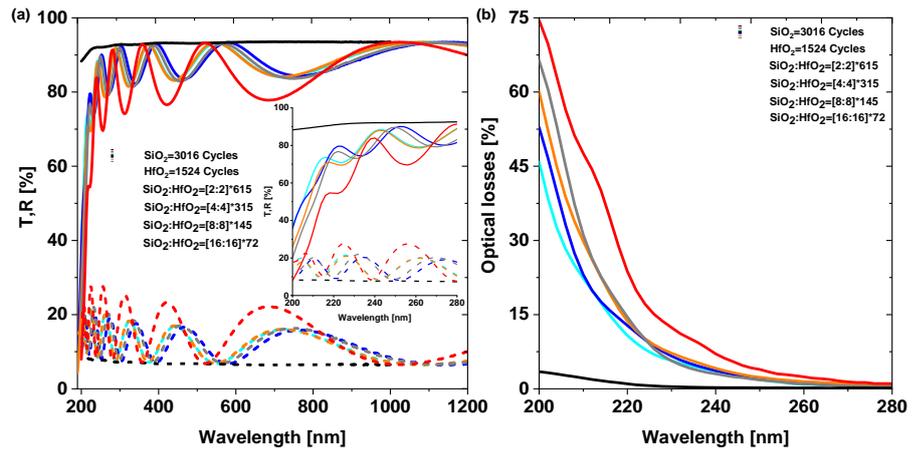

Fig. 6. (a) Transmittance T (solid curves), reflectance R (dashed curves), and (b) optical losses spectra for the $SiO_2$:$HfO_2$ heterostructures with identical cycle ratios of the constituents.

Additionally, the T and R spectra of the heterostructures with 4 and 8 cycles of $SiO_2$ were also measured in the VUV spectral range due to the shifting of absorption curves upon enhanced film properties. The spectra measured in the VUV range are graphically plotted along with the UV-VIS measurements in figure 7. Some spectral shifts could also be associated with the measurements at 2° and 6° angles of incidents for the VUV and UV-VIS ranges, respectively. Nevertheless, the spectra were consistent with the trend observed through both measurement ranges.

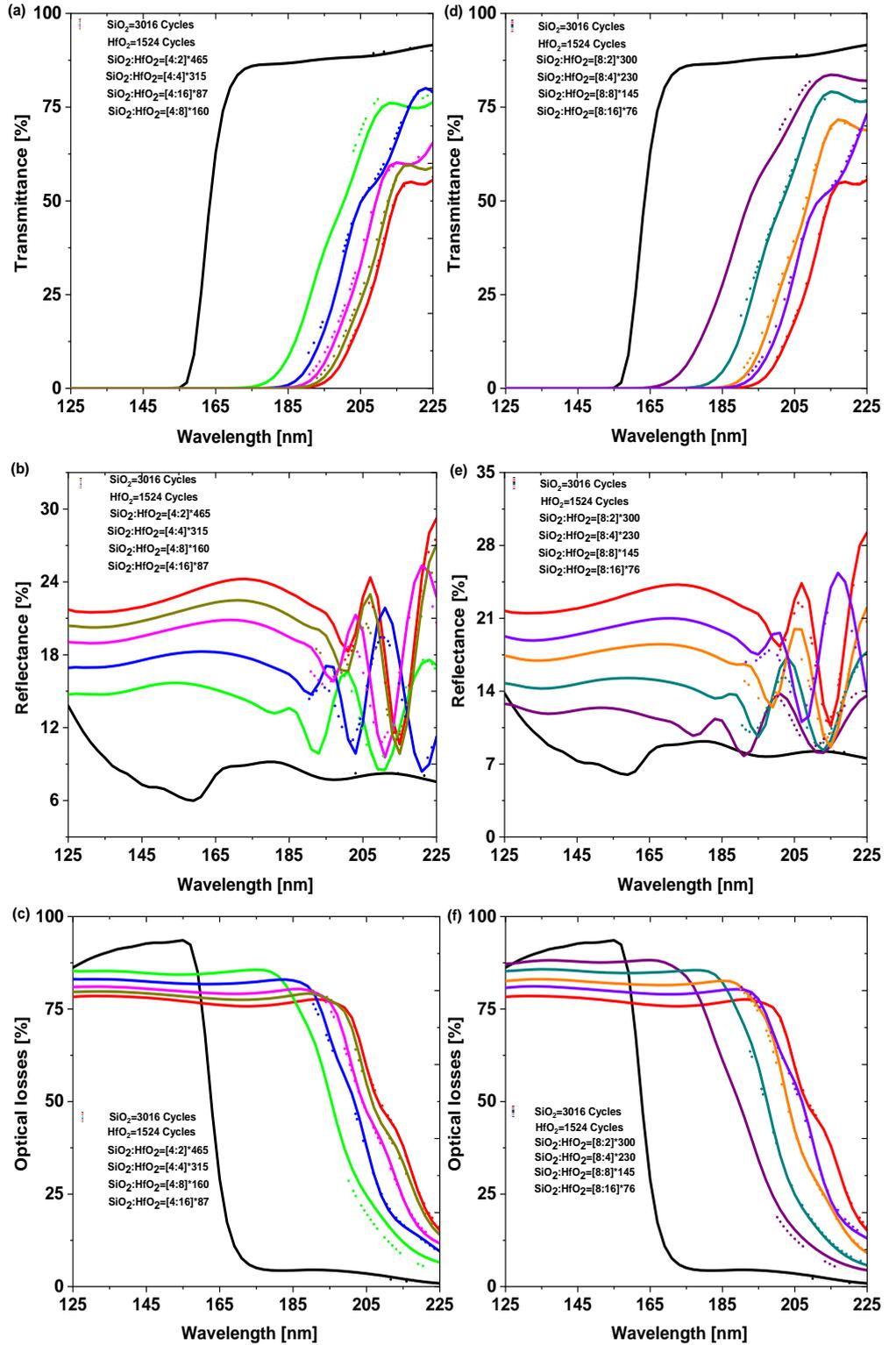

**Fig. 7**. Transmittance T, reflectance R, and optical losses spectra of SiO$_2$:HfO$_2$ heterostructures with (a-c) 4 and (d-f) 8 cycles of SiO$_2$ in the vacuum ultraviolet spectral ranges (dotted line UV-VIS, line VUV data).

Manipulating the mechanical stress of multilayer films is crucial for their optical performance. The mechanical strengths of the heterostructures are reported for 2, 4, 8, 16, 64 cycles of SiO$_2$ with increasing cycles of HfO$_2$. The residual stresses of such films were estimated by the wafer curvature measurement technique and graphically represented in figure 8. It was observed that the mechanical stress of the heterostructures can be controlled by adjusting the thickness of the ALD interlayers. The heterostructures with 2, 4, 8, and 16 cycles of SiO$_2$ exhibit tensile stress between 250 and 635 MPa. Heterostructures with identical cycle ratios of the constituents such as [2:2]*615, [4:4]*315, [8:8]*145, and [16:16]*72 reveal increasing stress due to the increase in interlayer thickness. However, heterostructures with 64 cycles of SiO₂ are quite complex and require further investigation as the films express tensile and compressive stresses. The residual stress for these heterostructures increases because of the gradual inclusion of more HfO₂ in the composition. However, the stress in the film should be optimized because the high coating stress might lead to functional degradation, especially for LIDT.

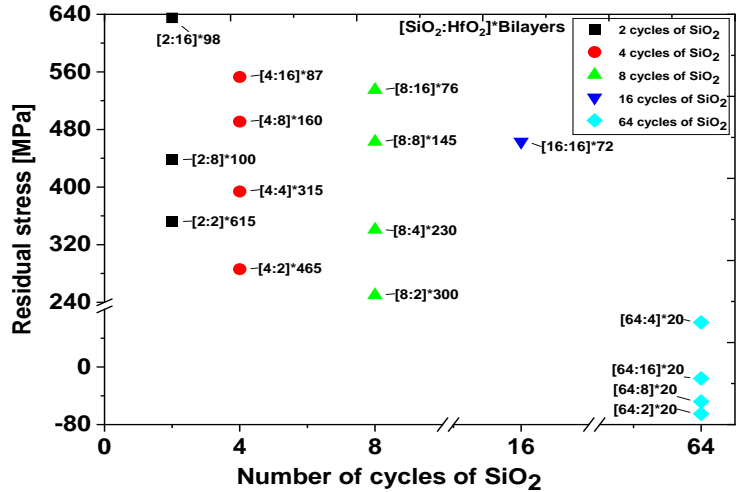

**Fig. 8.** Residual stress values of the SiO$_2$:HfO$_2$ heterostructure coatings.

*3.1.1. X-ray reflectivity analysis*

X-ray reflectivity (XRR) analysis of SiO$_2$:HfO$_2$ heterostructures was conducted on Si substrates at grazing angles (0° to 8°) to estimate the layer properties such as thickness, density, and surface roughness of the heterostructures. The measured and simulated curves of the SiO$_2$:HfO$_2$ films for a different number of ALD cycles of the constituent oxides denoting identical cycle ratios are presented in figure 9(a).

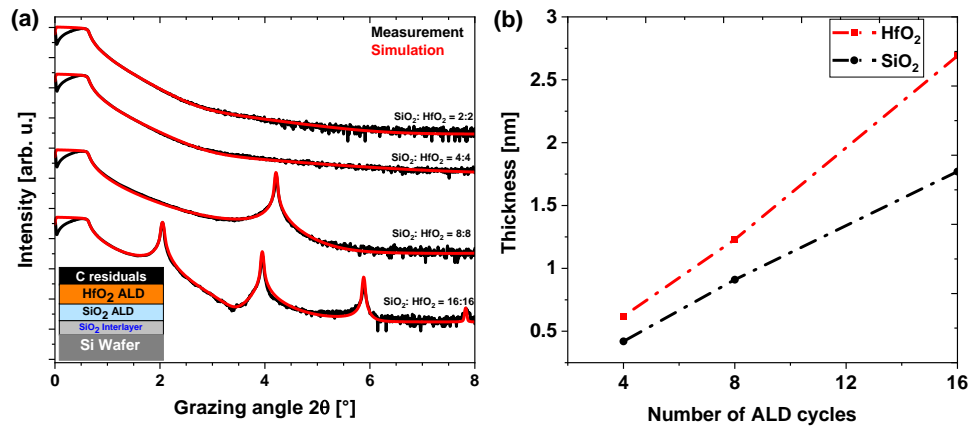

**Fig. 9.** (a) Measured and simulated reflectivity curves of SiO$_2$:HfO$_2$ heterostructures and (b) constituent layers thickness variation as a function of ALD cycles.

The XRR spectra were generated through the stack model consisting of the Si substrate, interfacial SiO$_2$ layer, SiO$_2$:HfO$_2$ heterostructures, and the C residuals on top (measured and simulated curves). The measured curves (black) show a very good agreement with the simulated spectra (red). The reflectivity curves show the presence of a sharp Bragg peaks, i.e., indicating superlattice characteristics, for the SiO$_2$:HfO$_2$ [8:8] and [16:16] heterostructures where the layer thickness for SiO$_2$ was 0.9 nm and 1.77 nm, respectively. The layer thickness of

HfO$_2$ for the heterostructures was recorded as 1.23 nm and 2.69 nm which was higher than that of SiO$_2$. It was observed that the layer thickness of the constituents increased almost two-fold when the number of ALD cycles doubled. ALD provides controlled growth of the heterostructures films having identical cycle ratios down to atomic scale. The periodic motifs for other heterostructures were not detected at the grazing incident angle because the peaks of the curves are disappearing due to the decrease in bilayer thickness. The layer thickness as a function of ALD cycles is depicted in figure 9 (b). The GPC of HfO$_2$ and SiO$_2$ can further be estimated from the slope of figure 9(b). The GPC of HfO$_2$ was higher than SiO$_2$ for the heterostructures due to the precursor properties. The total thickness of the heterostructures estimated by XRR was in good agreement with the spectroscopic ellipsometry analysis demonstrated in table 2. For the increasing number of ALD cycles, the density of HfO$_2$ film increased from 8.25 to 8.5 gcm$^{-3}$ compared to that of SiO$_2$ from 2.2 to 2.3 gcm$^{-3}$. The surface roughness of HfO$_2$ remained around 0.3-0.5 nm. Therefore, small surface roughness, as well as density of HfO$_2$, represents the amorphous state of the heterostructures because crystalline HfO$_2$ films exhibit high densities and rough surfaces due to columnar growth characteristics [19,21,85,86].

*3.1.2 Fourier transform infrared spectroscopy (FTIR) analysis*

The SiO$_2$: HfO$_2$ heterostructures along with single-layer SiO$_2$ and HfO$_2$ were investigated by the FTIR technique to identify the functional groups present in the films. The FTIR spectra were measured from 400 to 4500 cm$^{-1}$ ranges, however, presented in two different ranges of the spectrum in figure 10 (a,b). The different peaks in the spectra are identified as fingerprints of bonds and functional group regions of the films. The FTIR spectra were thickness normalized to make the peaks comparable for each film material. The normalized thickness was chosen to be 300 nm as most of the heterostructures were nearly in these thickness range. The FTIR peaks found in the curves were compared with the literature and denoted in figure 10.

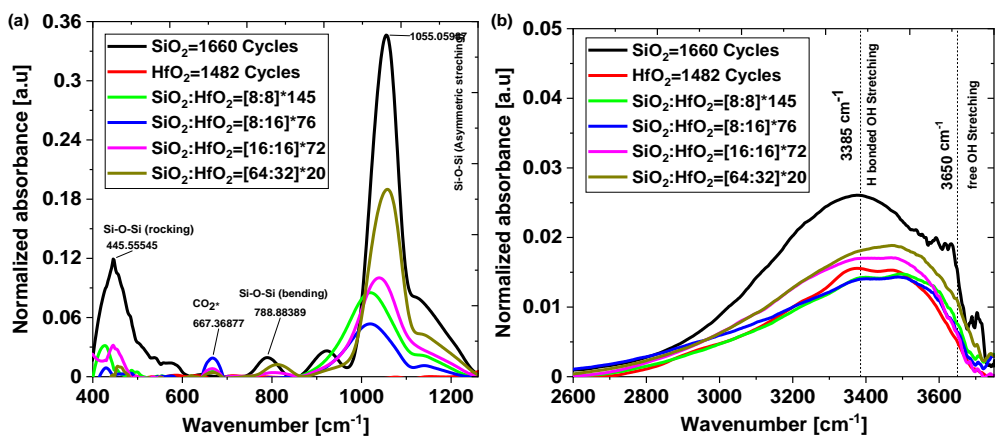

**Fig. 10**. FTIR absorption spectra of the heterostructures from (a) 400 to 1200 cm$^{-1}$ and (b) 2600 to 3750 cm$^{-1}$.

Three different vibrational modes of SiO$_2$ were detected at 446 cm$^{-1}$, 788 cm$^{-1}$, and 1055 cm$^{-1}$, respectively [60,87]. However, no peaks for HfO$_2$ films were encountered as the features of HfO$_2$ exceed our measurement ranges and the signal intensities were very low [53,87]. The SiO$_2$ films exhibit more OH groups compared to HfO$_2$ as shown in figure 10(b). Two major absorption peaks were detected for the heterostructures at 3385 and 3650 cm$^{-1}$ for different OH stretching modes [87,88]. The heterostructures were in good agreement with the findings of their constituents at 3650 cm$^{-1}$ as the film with significant influence of SiO$_2$ exhibits more OH groups, e.g., for heterostructures with 64 ALD cycles of SiO$_2$. The mechanisms of H bonded OH groups on the heterostructures are complex and not comparable with the inclusion of more SiO$_2$. Heterostructures with identical cycle ratios follow a similar trend at 3650 cm$^{-1}$ with a little deviation due to the different number of cycles SiO$_2$ and HfO$_2$. However, a vibrational mode of CO$_2$ was detected at 667 cm$^{-1}$ because the FTIR measurements were conducted at ambient conditions in the laboratory [87].

*3.1.2. Atomic force microscopy investigations*

To further understand the evolution of surface roughness on such coatings, AFM inspections were performed on selected compositions. Figure 11 shows the surface roughness values of the samples along with two extreme AFM micrographs. Single layer HfO2 film has the maximum r.m.s. roughness value of about 1.07 nm, whereas the heterostructure [8:2]*300 shows about 0.34 nm of r.m.s. surface roughness implying the fabrication of smooth multilayer stack. Further, the heterostructures with [2:2]*615 and [16:16]*72 indicate rms roughness of about 0.96 nm and 0.60 nm, respectively. In general, more SiO$_2$ content decreases the surface roughness. However, the number of interfaces may also lead to higher roughness values, as observed in the case of [2:2]*615 heterostructure. Thick SiO$_2$ have a low surface roughness and the quality of the ALD ultrathin coatings of silica is beneficial to reduce the surface roughness of the rougher hafnia [51,60].

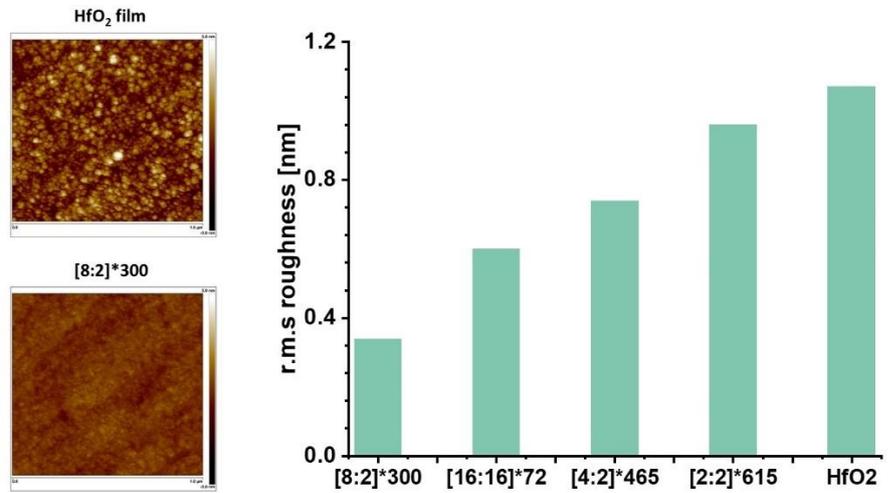

Fig. 11. Surface roughness (rms) values of SiO$_2$:HfO$_2$ heterostructures in comparison to single layer HfO$_2$ film. The corresponding AFM micrographs are included.

*3.1.3. Damage performance of SiO$_2$:HfO$_2$ heterostructures*

3.1.3.1 Nanosecond laser damage test at 355 nm wavelength

The SiO$_2$:HfO$_2$ heterostructure films grown on fused silica substrate were characterized by the irradiation of nanosecond laser pulses using the R-on-1 method because of the limited number of test sites available on the substrates (small substrate area). The LIDT test is governed by the test parameters which can influence the damage resistance of the coatings. Such parameters for Nd:YAG laser applied on R-on-1 test methods are specified in table 3.

**Table 3.** LIDT test parameters for nanosecond laser at 355 nm wavelength.

| Wavelength | Pulse length | Repetition rate | Incident angle | Spot diameter | Test method |
|---|---|---|---|---|---|
| 355 nm | 7 ns | 10 Hz | 10º | 200 to 300 μm | R-on-1 |

The damage test was initially conducted on the single-layer HfO$_2$ and SiO$_2$ films. SiO$_2$ coatings experienced damage immediately after the laser irradiation due to coatings errors. However, such films are well-known to exhibit high LIDT due to the wide bandgap [89–91]. The damage threshold of HfO$_2$ film was estimated at 16 J/cm$^2$ through regression analysis. Zhang et al. have reported an LIDT of 7 J/cm$^2$ [308 nm, 20 ns] grown by electron beam evaporation process [5].

They observed that HfO$_2$ deposited by ALD results in 31.8 J/cm$^2$ at 1064 nm wavelength with 3 ns pulses. Liu et al. have also reported an LIDT of 14 J/cm$^2$ [S-on-1, 1064 nm, 10 ns] for HfO$_2$ grown by ALD [50]. The LIDT was different in the UV and NIR spectral range due to the different laser conditioning parameters, test methods, and as well as deposition processes.[3,39,92,93]. LIDT directly scales with the wavelength, even if the remaining laser parameters, test methods and deposition process are kept intact. Additionally, pulse duration has a major influence on the damage mechanism. The damage threshold of the heterostructures was between 20 J/cm$^2$ and 79 J/cm$^2$ while increasing with the LIDT values. In the nanosecond pulse regime, the LIDT is influenced by the optical bandgap, i.e., the absorption edge. HfO$_2$ is the LIDT limiting material for the heterostructures in the UV spectral range because of its absorption edge. The LIDT as a function of the optical bandgap of the heterostructures is incorporated in figure 12(a). The heterostructures [8:2] *300 exhibit a relatively higher LIDT of about 79 J/cm$^2$ while incorporating more SiO$_2$ content in the heterostructure, hence ensuring a higher bandgap. The heterostructures with identical cycle ratios such as [2:2]*615 and [16:16]*72 demonstrate LIDT nearly in the similar range with slight deviations that are within the error margin. The heterostructure [2:16]*98 composed of higher relative content of HfO$_2$ reveals 20 J/cm$^2$ which are comparable with a single layer if the error margins are considered. Overall, the heterostructure film exhibits much higher LIDT than the single-layer films.

The LIDT in the nanosecond regime can also be explained in terms of refractive indices of SiO$_2$: HfO$_2$ heterostructures. The evolution of LIDT as a function of the tailored refractive indices obtained from heterostructures films are represented in figure 12(b), indicating an inverse relationship between them at 355 nm wavelength. The heterostructure [8:2]*300, having the highest optical bandgap and lowest surface roughness, exhibits the highest LIDT of 79 J/cm$^2$ (the lowest refractive index) due to the inclusion of more SiO$_2$ in the composition. An opposite scenario is noticeable for the [2:16]*98 heterostructure where the refractive indices are much higher due to the inclusion of more HfO$_2$. The heterostructures with identical cycle ratios are also consistent with such refractive indices. HfO$_2$ as a high index material reported the lowest LIDT of 16 J/cm$^2$ in this study.

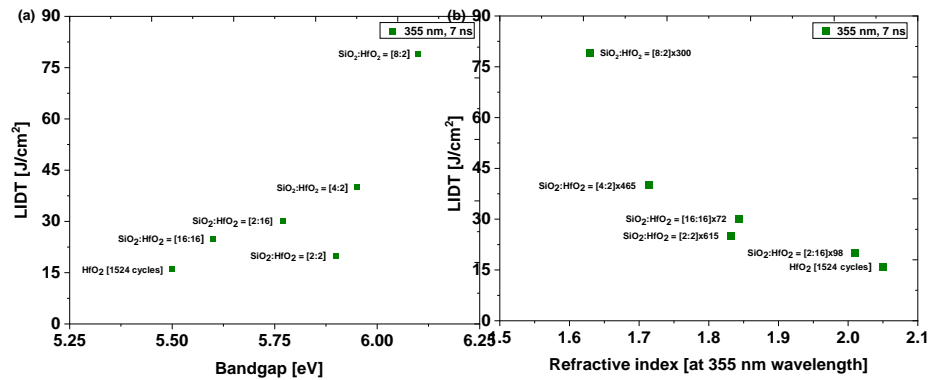

**Fig. 12.** LIDT vs. bandgap of various SiO$_2$: HfO$_2$ heterostructures, (b) LIDT vs. refractive indices of various SiO$_2$: HfO$_2$ heterostructures.

The damage resistance of the heterostructures also depends on the absorption because it may degrade the functionality of the films by introducing defects. The absorption measurements on the heterostructures were conducted using the photothermal common-path interferometry (PCI) method. The LIDT vs. absorption of selected heterostructures are depicted in figure 13. HfO$_2$ shows 278 ppm absorption at 355 nm wavelength which is significantly higher than that of SiO$_2$ about 4.5 ppm; as HfO$_2$ is an absorbing material in the UV spectral range. The heterostructures with more incorporation of HfO$_2$ experience high absorption and comparatively low LIDT such as the [2:16]*98 heterostructure (82 ppm, 20 J/cm$^2$). The scenario is exactly the opposite for the [8:2]*300 and [4:2]*465 heterostructures where the LIDT is much higher with low absorption (addition of more SiO$_2$). The heterostructures with identical cycle ratios were also consistent with such properties of the heterostructures.

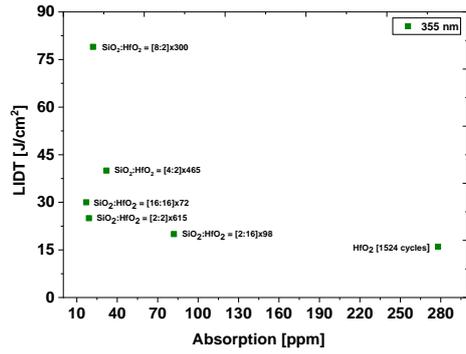

**Fig. 13.** The effect of absorption on LIDT of selected heterostructures.

*3.2 Antireflection Coatings*

The antireflection (AR) coatings designed at 355 nm wavelength will be reported in this section because the heterostructures incorporate $SiO_2$ and $HfO_2$ layers that exhibit low optical absorption in the UV and NIR spectral range. The multilayer stack of homogeneous and inhomogeneous coatings will also be demonstrated. Inhomogeneous coatings are also known as the graded-index or rugate coatings where refractive indices are varied along with the interlayer thickness which can be obtained by adjusting the constituents cycle ratio of the heterostructures. The coatings were designed using the software Optilayer, version 12.83g (OptiLayer GmbH, Garching, Germany), whereby $HfO_2$ and $SiO_2$ were used as the high and low refractive index material, respectively. The coatings were optimized between 0° and 45° angle of incidence.

3.2.1 Antireflection coatings at 355 nm wavelength

Here, we report antireflection coatings designed for 355 nm wavelength using conventional multilayer and graded index structures grown on fused silica substrates. The design pattern for a multilayer coatings stack is given in table 5. The four-layer coatings start with a $HfO_2$ layer on the substrate followed by two alternating layers and ends with a $SiO_2$ layer. The corresponding refractive index profile of the design is illustrated in figure 14. The coatings were designed for a total thickness of 152.3 nm.

**Table 5.** Design of multilayer stack antireflection coatings at 355 nm wavelength.

| Layer | Material | Physical thickness (nm) | Optical thickness (nm) | Refractive index |
|---|---|---|---|---|
| 1 | $HfO_2$ | 13.5 | 26.1 | 1.93 |
| 2 | $SiO_2$ | 23.8 | 34.6 | 1.45 |
| 3 | $HfO_2$ | 48.6 | 93.9 | 1.93 |
| 4 | $SiO_2$ | 66.4 | 96.4 | 1.45 |
|   |   | Total = 152.3 |   |   |

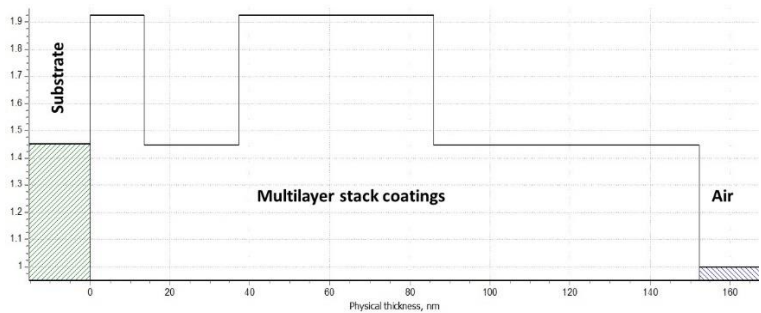

**Fig. 14.** Refractive index profile of multilayer stack antireflection coatings at 355 nm.

The refractive indices of the intermediate layers vary due to the distinct composition cycle ratio of $SiO_2$ and $HfO_2$. In comparison to the multilayer stack, the graded coatings incorporate interlayers between the discontinuous interfaces. The graded AR design was optimized for a total thickness of 149.1 nm and is given in table 6. The corresponding refractive index profile of the design is given in figure 15.

The curves of the designed and measured reflectance for the two types of antireflection coatings are shown in figure 16. The coatings on the double-sides (ds) of the fused silica substrates were designed and prepared at a center wavelength of 355 nm for 6° and 45º angle of incidence (AOI). In the case of multilayer stacks, the AR films on the ds-coated substrates were designed to achieve reflectance less than 0.35% at 355 nm wavelength (6° AOI). The measured spectra of the AR coatings were in good agreement with the designed reflectance minima. A similar scenario was observed for the reflectivity curves at 45° AOI. The graded-index layer coatings were designed to achieve reflectance less than 0.4% and 1% at 355 nm wavelength for 6° and 45° AOI, respectively. However, the shift of measured reflectance minima of the AR coatings at 45° AOI towards a higher wavelength range could be attributed to some thickness mismatch with the design.

**Table 6.** Design of graded-index antireflection coatings at 355 nm wavelength.

| Layer | Material | Composition | Physical thickness (nm) | Fraction of $HfO_2$ (%) | Refractive index | Number of cycles/bilayers |
|---|---|---|---|---|---|---|
| 1 | $HfO_2$ | Single layer | 4.7 | 100 | 1.93 | 29 |
| 2 | $SiO_2$:$HfO_2$ | 4:8 | 4.7 | 82.57 | 1.84 | 3 |
| 3 | $SiO_2$:$HfO_2$ | 8:8 | 4.7 | 51.71 | 1.70 | 2 |
| 4 | $SiO_2$:$HfO_2$ | 8:2 | 4.7 | 18.23 | 1.54 | 4 |
| 5 | $SiO_2$ | Single layer | 9.3 | 0 | 1.45 | 80 |
| 6 | $SiO_2$:$HfO_2$ | 8:2 | 4.7 | 17.78 | 1.53 | 4 |
| 7 | $SiO_2$:$HfO_2$ | 8:8 | 4.7 | 48.43 | 1.68 | 2 |
| 8 | $SiO_2$:$HfO_2$ | 4:8 | 4.7 | 78.27 | 1.82 | 4 |
| 9 | $SiO_2$:$HfO_2$ | 2:8 | 4.7 | 96.31 | 1.91 | 3 |
| 10 | $HfO_2$ | Single layer | 32.6 | 100 | 1.93 | 201 |
| 11 | $SiO_2$:$HfO_2$ | 4:8 | 4.7 | 76.04 | 1.81 | 3 |
| 12 | $SiO_2$:$HfO_2$ | 8:4 | 4.7 | 35.63 | 1.62 | 3 |
| 13 | $SiO_2$:$HfO_2$ | 64:2 | 4.7 | 4.26 | 1.47 | 1 |
| 14 | $SiO_2$ | Single layer | 55.9 | 0 | 1.45 | 482 |
| | | | Total=149.1 | | | |

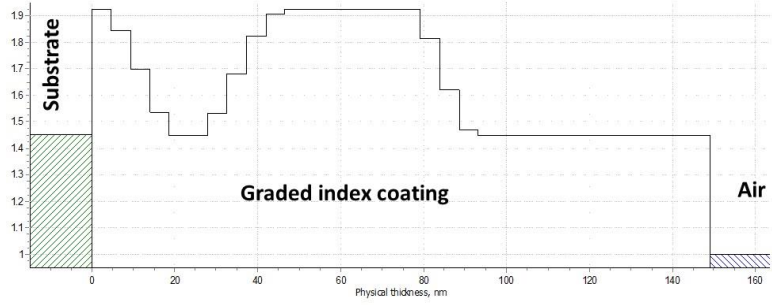

**Fig. 15.** Refractive index profile of graded-index antireflection coatings at 355 nm.

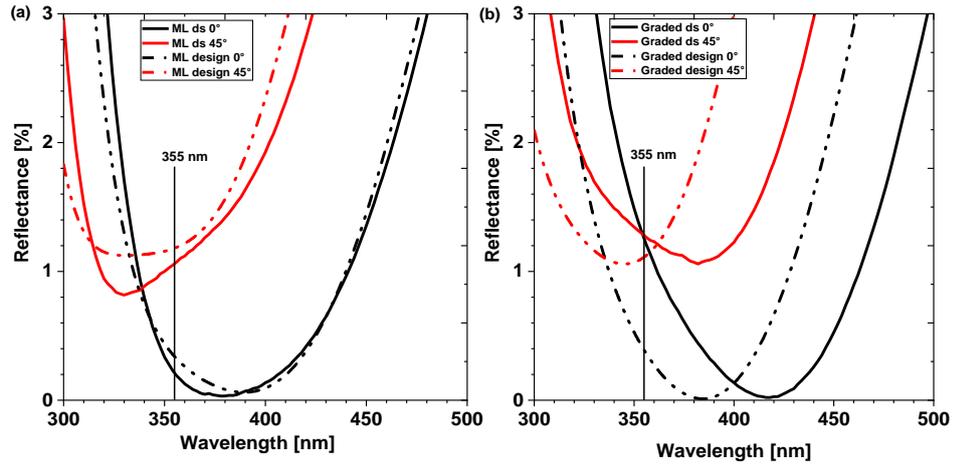

**Fig. 16.** Comparison of designed and measured reflectance spectra for (a) multilayer stack and (b) graded antireflection coatings at 355 nm wavelength.

The laser damage resistance tests of the antireflection coatings at 355 nm wavelength were conducted using nanosecond laser pulses with a similar conditioning parameter of the R-on-1 method as stated in table 3. The LIDT of the AR coatings exhibits 29±6 J/cm² and 31±7 J/cm² for multilayer stack and graded structure, respectively, as mentioned in table 7. The LIDT of graded and multilayer stack coatings are comparable and differ by about 2 J/cm² which are within the measurement errors. Chaoyi et al. have reported an LIDT of 24.4 J/cm² in the UV regime (355 nm wavelength) for $HfO_2:SiO_2$ films grown by the PEALD technique [1]. The laser damage threshold of graded-index coatings could have been much higher because of its robust design. Therefore, damage initiation mechanisms in the UV regime for nanosecond laser pulses might be linked to absorbing defects in the coatings [94,95]. The absorption measurements on the AR coatings were carried out at 355 nm wavelength by using the photothermal common-path interferometry (PCI) and laser-induced deflection (LID) methods. The absorption values obtained from these techniques are summarized in table 7.

**Table 7.** Absorption measurements for AR coatings at 355 nm wavelength.

| Coating structure | Method: PCI (unannealed) | Method: LID (unannealed) | Method: LID (annealed) | LIDT |
|---|---|---|---|---|
| Multilayer stack | 16.7 ± 5.1 [ppm] | 86.5 ± 2.2 [ppm] | 26.5 ± 3.5 [ppm] | 29 ± 6 [J/cm²] |
| Rugate coating | 10.2 ± 3.2 [ppm] | 85.3 ± 8.0 [ppm] | 58.0 ± 7.5 [ppm] | 31 ± 7 [J/cm²] |

The measurement using the PCI technique provides a difference of 6.5 ppm between the graded and multilayer stack which can be even less because fused silica substrate contributes 1-2 ppm absorption [50]. This method considered the coatings that did not undergo annealing at a higher temperature. In the case of LID, the measurements were conducted at the same wavelength for both annealed and unannealed coatings which provide significant differences in absorptions. For the unannealed coatings, the absorption values are comparable with graded and multilayer stacks except for the errors. This is not the case for the annealed coatings since the absorption for rugate (58±7.5 ppm) is approximately double compared to the multilayer stack (26.5±3.5 ppm). The absorption for both coatings decreased significantly because of annealing which indicates the presence of hydroxyl groups in the thin film coatings. Therefore, annealing reveals a substantial reduction of weak absorption in the coatings and can increase the LIDT. The LIDT estimated in our case is applicable for the unannealed coatings only. However, the annealing at a high temperature can also transform the amorphous phase of the coating materials like $HfO_2$ into a crystalline state. That absorption values determined using PCI and LID techniques show quite difference which could be a scope of further investigations. The measurement threshold ranges for these techniques might also be different owing to the asymmetrical calibration methods of the systems.

## 4. Conclusion

In this work, we carried out a comprehensive study on the optical properties of $SiO_2$:$HfO_2$ heterostructures developed by the PEALD method for high power laser optics. It was observed that the refractive indices, extinction coefficients, absorption edges, the residual stress, and the optical bandgaps of the heterostructure films are tunable by precise tuning of compositions. Extended transparent ranges and intermediate refractive indices of the heterostructures provide flexibility in thin film layer design. The optical bandgaps of the materials show an inverse relationship with the refractive indices at the reported wavelengths. FTIR studies on the heterostructures reveal more OH groups are present in $HfO_2$ than $SiO_2$. The XRR studies reveal that the heterostructures were in an amorphous state facilitating low loss optical coatings. The small surface roughness of the heterostructures indicates that the coatings were very smooth and uniform in thickness. The LIDT tests of the heterostructures were carried out using nanosecond pulses by employing the R-on-1 method. The LIDT of the heterostructures was in good agreement with the literature for nanosecond pulses using the R-on-1 test method. The LIDT of the heterostructures in the nanosecond pulse regime can be scaled with their bandgap and refractive indices. The absorption played an additional role in the damage estimation of such coatings because the LIDT was decreased significantly with increased absorption losses. The absorption in the $HfO_2$ films was higher in the UV regime because of its absorption edges. Two types of antireflection coatings (multilayer stacks and graded-index structures) were developed for high power laser applications in the UV spectral range. The graded index coatings were designed to reduce the effects of discontinuous interfaces (by incorporating intermediate layers) in the coatings. However, graded-index coatings could not help in enhancing the LIDT of the heterostructures, which could be further investigated. Several factors such as conditioning of laser parameters and shifting of electric field distributions to the low index material from the layer interfaces (vulnerable to the incident laser intensity) through sophisticated design will also be taken into consideration. High reflection coatings ($\lambda$/4 optical thickness) and functionalization of 3D optical surfaces can be developed by ALD and their LIDT will be examined in the future. Overall, this work demonstrates a possible route via PEALD to develop atomically thin heterostructures of $SiO_2$:$HfO_2$ with tailored optical properties.

**Author Contributions:** For research articles with several authors, a short paragraph specifying their individual contributions must be provided. The following statements should be used "Conceptualization, A.S.; methodology, S.A., P.P., V.B., A.S.; formal analysis, S.A, P.P., V.B., O.S., M.T., G.M., S.R., F.O., A.G.; investigation, S.A., P.P.; writing—original draft preparation, S.A.; writing—review and editing, S.A., P.P., M.T., O.S., S.W., S.S., S.R., T.F., A.S; supervision, A.S.; project administration, A.S. and S.N.; funding acquisition, A.S., S.N. All authors have read and agreed to the published version of the manuscript.

**Funding**
This research was financially supported by Deutsche Forschungsgemeinschaft (DFG, German Research Foundation) Project-ID 398816777–SFB 1375, the BMWi AiF Project-ID ZF4309604SY8 and the Fraunhofer Society Attract 066-601020.

**Acknowledgements**
The authors gratefully acknowledge Paul Schmitt for the XRR measurements and David Kästner for the technical support.

**Conflicts of interest**
The authors declare no conflicts of interest.